%% file: mMIMO_indoor.tex
\documentclass[final]{IEEEtran}

\newlength \figwidth
\usepackage{cite}
\ifCLASSINFOpdf
  \usepackage[pdftex]{graphicx}
	\usepackage{epstopdf}
  \graphicspath{{../Figures/}}
  \DeclareGraphicsExtensions{.eps}
\else
  \usepackage[dvips]{graphicx}
\fi
\usepackage[cmex10]{amsmath}
\usepackage{amssymb}
\usepackage{amsthm}
\usepackage{url}
\usepackage{dsfont}
\usepackage{tabulary}
\usepackage{multirow}
\usepackage{color}

\usepackage{color}
\usepackage{times}
\usepackage{verbatim}
\usepackage{floatflt}
\usepackage{enumerate}
\usepackage{array}
\usepackage{multicol,afterpage,wrapfig} 
\usepackage{tikz}
\usepackage{algorithm}
\usepackage[noend]{algpseudocode}
\makeatletter
\def\BState{\State\hskip-\ALG@thistlm}
\makeatother
\usepackage{amsmath,amssymb,eucal}
\usepackage[utf8]{inputenc}
\usepackage{epsfig}
\usepackage{exscale}
\usepackage{latexsym}
\usepackage{verbatim}
\usepackage{amsfonts}
\usepackage{multirow}
\usepackage{cite}
\usepackage{balance}
\usepackage{color}
\usepackage{transparent}
\usepackage{import}
\usepackage{mathtools}
\usepackage{tikz}
\usetikzlibrary{automata,arrows,positioning,calc}
\usepackage{bbm}
\usepackage[printonlyused,withpage]{acronym}
\usepackage{tabu}
\usepackage{longtable}
\usepackage{balance}
\usepackage{footnote}
\usepackage{tablefootnote}
\usepackage[font=footnotesize]{caption}
\usepackage{enumitem}

\def\BibTeX{{\rm B\kern-.05em{\sc i\kern-.025em b}\kern-.08em
    T\kern-.1667em\lower.7ex\hbox{E}\kern-.125emX}}
    
\renewcommand{\IEEEQED}{\IEEEQEDopen}

\newcommand*\xbar[1]{%
  \hbox{%
    \vbox{%
      \hrule height 0.5pt % The actual bar
      \kern0.36ex%         % Distance between bar and symbol
      \hbox{%
        \kern-0.12em%      % Shortening on the left side
        \ensuremath{#1}%
        \kern-0.12em%      % Shortening on the right side
      }%
    }%
  }%
}

\input{macros}

\usetikzlibrary{arrows,calc}
\allowdisplaybreaks

\IEEEoverridecommandlockouts
\begin{document}
\pagenumbering{gobble}
%\makeatletter
%\def\thm@space@setup{\thm@preskip=0pt
%\thm@postskip=0pt}
%\makeatother

\newtheorem{Theorem}{\bf Theorem}
\newtheorem{Corollary}{\bf Corollary}
\newtheorem{Remark}{\bf Remark}
\newtheorem{Lemma}{\bf Lemma}
\newtheorem{Proposition}{\bf Proposition}
\newtheorem{Assumption}{\bf Assumption}
\newtheorem{Definition}{\bf Definition}
\title{Indoor Massive MIMO Deployments for\\ Uniformly High Wireless Capacity}
\author{\IEEEauthorblockN{{Giovanni~Geraci$^{\star}$, Adrian~Garcia-Rodriguez$^{\star}$, David~L\'{o}pez-P\'{e}rez$^{\star}$, \\ Lorenzo~Galati~Giordano$^{\star}$, Paolo~Baracca$^{\diamond}$, and Holger~Claussen$^{\star}$}}\\
\normalsize\IEEEauthorblockA{\emph{Nokia Bell Labs, $^{\star}$Ireland and $^{\diamond}$Germany}}}
\maketitle
\thispagestyle{empty}

\IEEEpeerreviewmaketitle
\input{Abstract}
\input{Section1}
\input{Section2}
\input{Section3}
\input{Section4}

\input{Section5}
\ifCLASSOPTIONcaptionsoff
  \newpage
\fi
\bibliographystyle{IEEEtran}
\bibliography{Strings_Gio,Bib_Gio}
\end{document}

%% file: macros.tex
\newfont{\bbb}{msbm10 scaled 500}

\newfont{\bb}{msbm10 scaled 1100}

\newcommand{\executeiffilenewer}[3]{%
\ifnum\pdfstrcmp{\pdffilemoddate{#1}}%
{\pdffilemoddate{#2}}>0%
{\immediate\write18{#3}}\fi%
}

%% file: Abstract.tex
\begin{abstract}
%\vspace*{4cm}
Providing consistently high wireless capacity is becoming increasingly important to support the applications required by future digital enterprises. In this paper, we propose \textit{Eigen-direction-aware ZF (EDA-ZF)} with partial coordination among base stations (BSs) and distributed interference suppression as a practical approach to achieve this objective. We compare our solution with
\textit{Zero Forcing (ZF)}, entailing neither BS coordination or inter-cell interference mitigation, and
\textit{Network MIMO (NeMIMO)}, where full BS coordination enables centralized inter-cell interference management.
We also evaluate the performance of said schemes for three sub-6 GHz deployments with varying BS densities -- sparse, intermediate, and dense -- all with fixed total number of antennas and radiated power. 
%: \emph{sparse} -- two 64-antenna BSs, each radiating 24 dBm; \emph{intermediate} -- eight 16-antenna BSs, each radiating 18 dBm; and \emph{dense} -- 32 four-antenna BSs, each radiating 12 dBm.
Extensive simulations show that:
(i) \emph{indoor massive MIMO} implementing the proposed EDA-ZF provides uniformly good rates for all users; 
%, without requiring full coordination.
(ii) indoor network densification is detrimental unless full coordination is implemented;
(iii) deploying NeMIMO pays off under strong outdoor interference, especially for cell-edge users.%, at the expense of introducing significant backhaul loads.
\end{abstract}

%% file: Section1.tex
\section{Introduction}

%\cite{DieZirJag2104}
%\cite{HavZim2017,Grier2017,PorHep2014}. 

The fourth industrial revolution underway -- \emph{Industry 4.0} -- urgently demands fast, cable-less, and reliable exchange of information between sensors, humans, and smart machines, often found indoors and in large numbers. It is a timely and critical task to guarantee said ubiquitous in-building wireless connectivity for enterprises and public institutions alike \cite{WolSauJas2017}. While spectrally efficient multi-antenna cellular systems are regarded as the best candidate to meet this demand, they are inherently interference limited \cite{AndChoHea2007}, and they may lose much of their effectiveness in densely populated indoor scenarios. Handling indoor interference may thus be identified as the ultimate task to achieve an ICT-enabled smart industry.

\subsection{Related Work and Contribution}

In this paper, we consider three interference management schemes for multi-antenna indoor deployments:
\begin{itemize}
\item \textit{Zero Forcing (ZF)} -- without base station (BS) coordination nor inter-cell interference management;
\item \textit{Network MIMO (NeMIMO)} -- where full BS coordination enables centralized inter-cell interference management;
\item Novel \textit{Eigen-direction-aware ZF (EDA-ZF)} -- with partial BS coordination and distributed interference suppression.
\end{itemize}

With conventional \textit{ZF} \cite{SpeSwiHaa:04}, each BS simultaneously serves its scheduled users (UEs) via spatial multiplexing, suppressing all intra-cell crosstalk. In spite of this, the lack of inter-cell interference management results in poor user rates, especially for those UEs located at the cell edge. %While ZF does not require any coordination among BS, this comes at the expense of the inability to perform inter-cell interference management, resulting in poor rates especially at the cell edge. 

With \textit{NeMIMO} \cite{VenLozVal2007,HosYuAdv2016} -- also known in the literature as 
%\cite{ZhaCheAnd2009}
cooperative multipoint (CoMP) \cite{LeeSeoCle2012}, %\cite{GesHanHua2010,NigMinHae:14,XuYanLi2014}
distributed MIMO \cite{BalRogMic2013}, %\cite{HamRahAdb2016,NgEvaHan2008}, 
cell-free MIMO \cite{NgoAshYan2017}, and pCell \cite{artemis2015} %\cite{artemis2015,ForPerSai2015}
-- all BSs cooperate to jointly serve all UEs, boosting the cell-edge user throughput. This requires sharing information about the set of scheduled UEs, their channel training resources, and -- more importantly -- all data intended for all scheduled UEs, e.g., through a wired backhaul. Moreover, NeMIMO also requires a tight symbol-level synchronization among BSs, which complicates its practical implementation \cite{R117011313,LozHeaAnd:13}.

In this paper, we propose \textit{EDA-ZF} as a more practical alternative to improve performance at the cell-edge users through distributed interference mitigation. With EDA-ZF, BSs steer the inter-cell interference towards the nullspace of neighboring UEs. In order to do so, BSs are required to share scheduling and pilot allocation information, but -- unlike NeMIMO -- no user data information.

While recent attempts to distributed interference mitigation have been made in \cite{HoyHosTen2014,BjoLarDeb2016,YanGerQueTSP2016,GerGarLop2016,GarGerGal2017,GarGerLop2017GC}, 
%\cite{ZakHan:12,HoyHosTen2014,BjoLarDeb2016,ZhuWanQia:16,YanGerQueTSP2016,GerGarLop2016,GarGerGal2017,GarGerLop2017GC}, 
the current paper and the proposed EDA-ZF approach differ from these works in a number of key aspects: 
unlike \cite{HoyHosTen2014}, EDA-ZF employs inter-cell channel state information (CSI) to place radiation nulls, rather than to regularize the precoder to mitigate inter-tier interference; 
unlike \cite{BjoLarDeb2016,YanGerQueTSP2016}, EDA-ZF targets the eigen-directions of the most vulnerable UEs, rather than all neighboring UEs; furthermore, unlike \cite{HoyHosTen2014, BjoLarDeb2016,YanGerQueTSP2016}, this paper considers an indoor deployment, which exhibits considerably different features due to the large number of interfering line-of-sight (LoS) links; 
finally, unlike \cite{GerGarLop2016,GarGerGal2017,GarGerLop2017GC}, this paper focuses on a cellular architecture operating in a licensed band.

\subsection{Approach and Summary of Results}

We evaluate the performance of the three interference management schemes -- ZF, NeMIMO, and EDA-ZF -- in three different sub-6 GHz indoor deployment scenarios. In all three scenarios, the total number of antennas and the total radiated power are kept fixed, in order to perform a fair comparison:
\begin{itemize}
\item A \emph{sparse} deployment of two 64-antenna massive MIMO BSs, each radiating 24 dBm.
\item An \emph{intermediate} deployment of eight 16-antenna BSs, each radiating 18 dBm.
\item A \emph{dense} deployment of 32 four-antenna small cell BSs, each radiating 12 dBm.
\end{itemize}
A number of key conclusions can be drawn from our study:
\begin{itemize}
\item A sparse deployment implementing the proposed scheme – \emph{EDA-ZF indoor massive MIMO} – provides uniformly good performance for all UEs. In particular, the achievable rates are very close to the ones attained by NeMIMO, without requiring full coordination among BSs.
\item Due to the strong indoor LoS interference, network densification is detrimental unless full coordination (NeMIMO) is implemented.
\item In the presence of strong outdoor co-channel interference, a dense, fully coordinated deployment -- NeMIMO small cells -- rewards cell-edge UEs, but at the expense of significant backhaul synchronization requirements.
\end{itemize} 

%% file: Section2.tex
%\vspace*{-0.3cm}
\section{System Setup}

\subsection{Deployment}

\begin{figure}[!t]
    \centering
        \includegraphics[width=0.9\columnwidth]{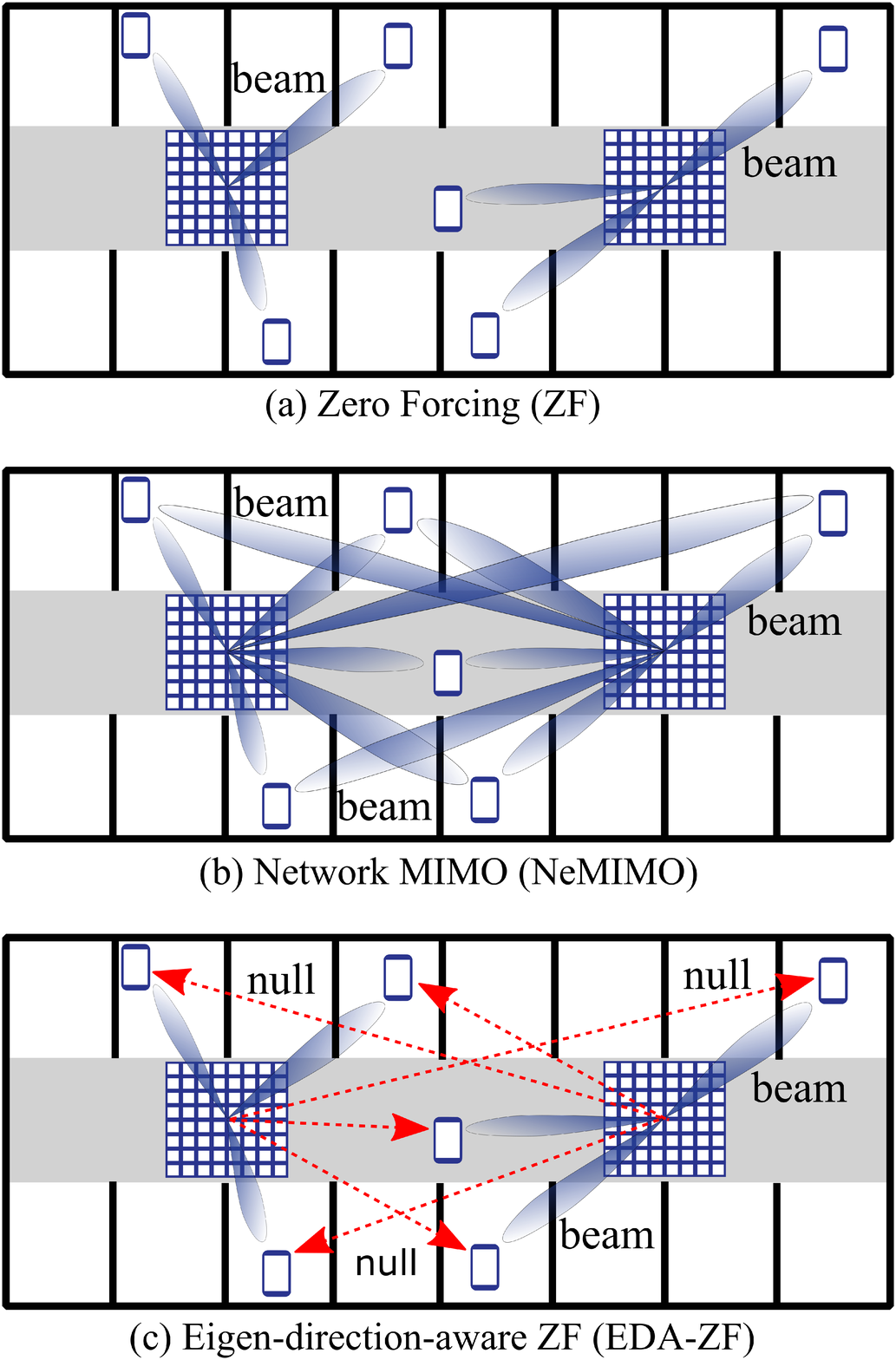}
    \caption{Illustration of the three interference management schemes, for $N_{\textrm{B}}=2$ BSs, $N_{\textrm{A}}=64$ antennas per BS, and $N_{\textrm{U},b}=3$ scheduled UEs per BS. (a) ZF: no inter-cell interference management; (b) NeMIMO: all BSs jointly serve the UEs scheduled in all cells, suppressing all interference; (c) EDA-ZF: each BS serves its own set of UEs, while placing $N_{\textrm{N}}=3$ radiation nulls towards UEs in neighboring cells.}
		\label{fig:threeSchemes}
\end{figure}

We consider the single-floor $120~\textrm{m}\times 50~\textrm{m}$ indoor hotspot network depicted in Fig.~\ref{fig:threeSchemes}. In this setting, which is conventionally recommended for indoor studies \cite{3GPP38802}, an operator deploys a certain number of BSs $N_{\textrm{B}}$ on the ceiling to complement its outdoor network and enhance user capacity.  Let $\mathcal{B}$ denote the set of deployed BSs, which comply to an individual maximum transmit power constraint $P_{\mathrm{B}}$. We assume that UEs associate to the BS that provides the largest average received signal strength (RSS) and that each BS $b \in \mathcal{B}$ schedules a subset of its associated UEs for transmission \cite{LopDinCla:15}. The set of scheduled UEs on a given time-frequency physical resource block (PRB) and its cardinality are denoted by $\mathcal{U}_b$ and $N_{\mathrm{U},b}$, respectively. In what follows, we will denote by
\begin{equation}
\mathcal{U}  = \underset{b \in \mathcal{B}}{\bigcup} \mathcal{U}_b	
\end{equation}
the set of UEs scheduled by all BSs on a given PRB, and by
\begin{equation}
N_{\mathrm{U}} = \sum_{b \in \mathcal{B}}{N_{\mathrm{U},b}}
\end{equation}
the cardinality of said set $\mathcal{U}$.
 %the maximum number of UE that any BS can schedule on a given PRB, i.e., $N_{\mathrm{U},b} \leq N_{\mathrm{U}}$, $\forall b$.
 %to guarantee a full coverage, i.e., a minimum received signal strength (RSS) of -82 dBm for all users (UEs) located across the floor.
 
\subsection{Channel Model}

The considered indoor setup constitutes a challenging deployment due to the physical proximity between nodes. This is because the interference experienced by nodes reusing the same PRB is significantly larger than that perceived in more sparse outdoor deployments. Indeed, the probability of LoS $P_{\mathrm{LoS}}$ as a function of the 3D distance $d$ in meters between any two nodes follows \cite{3GPP36889}
\begin{equation}
  P_{\mathrm{LoS}} = \left\{ 
  \begin{array}{c c l}
  1 & \quad \textrm{if} \enspace d \leq 18\\
	e^{-\frac{d-18}{27}} & \quad \textrm{if} \enspace 18 < d \leq 37\\
  0.5 & \quad \textrm{if} \enspace d > 37.\\
	\end{array} \right.
\label{eqn:PLOS}
\end{equation}

In the following we consider that propagation channels are affected by slow channel gain (comprising antenna gain, path loss, and shadowing) and fast fading \cite{3GPP38802}. We adopt a block-fading propagation model, and assume channel reciprocity since uplink/downlink (UL/DL) transmissions share the same frequency band through time division duplexing (TDD). We consider that all UEs are equipped with a single antenna, and that each BS is comprised of $N_{\mathrm{A}}$ antennas. We also let $\mathbf{h}_{ib} \in \mathbb{C}^{N_{\mathrm{A}}}$ denote the channel vector between UE $i$ and BS $b$. In this setup, each BS $b \in \mathcal{B}$ can obtain an estimate of the channel $\mathbf{h}_{ib}$ to/from each scheduled UE $i \in \mathcal{U}_b$ via pilot signals transmitted by the UE during a training phase \cite{Mar:10}. In this paper, we assume that all UEs convey orthogonal pilot signals.\footnote{Employing orthogonal pilots is essential in the indoor scenario considered, where severe pilot contamination due to the high user density would make data transmission through spatial multiplexing infeasible.} % Studies considering additional impairments are left for future work. %such estimate is considered to be perfect %In this paper, such estimate is assumed to be perfect for all three interference management schemes under study, thus not altering their comparison. 
Following \cite{NguWigKov2017}, we also account for the presence of outdoor interference, which may be generated, e.g., by a co-channel macro BS pointing towards the building of interest.

%The signal $\mathbf{z}_{i}$ received by UE $i$ on a certain PRB is given by
%\begin{align}
%\mathbf{z}_{i} = \! \! \sum_{j \in \mathcal{A}^{\star} \cup \mathcal{U}^{\star}} \!\! \sqrt{P_{j}} \, \mathbf{H}_{ij}^{\mathrm{H}} \mathbf{W}_{j} \mathbf{s}_{j} \!+\! \boldsymbol{\epsilon}_{i},
%\label{eqn:zi}
%\end{align}
%where:
%\begin{itemize}
%\item $\mathbf{H}_{ij} \in \mathbb{C}^{M_{i}\times M_{j}}$ denotes the channel matrix between nodes $j$ and $i$,
%\item $\mathbf{W}_{j} \in \mathbb{C}^{M_{j}\times K_{j}}$ is the normalized precoding matrix employed by node $j$, 
%\item $\mathbf{s}_{j} \in \mathbb{C}^{M_{j}}$ is the unit-variance signal vector, 
%\item $P_j$ denotes the average transmission power of node $j$, and
%\item $\boldsymbol{\epsilon}_{i} \in \mathbb{C}^{M_{i}}$ is zero-mean complex Gaussian thermal noise with variance $\sigma^2_{\epsilon}$.
%\end{itemize}

%% file: Section3.tex
\section{Interference Management Schemes}

We now detail the three interference management schemes considered in this paper and shown in Fig.~\ref{fig:threeSchemes}, namely: (a) ZF, (b) NeMIMO, and (c) EDA-ZF. We concentrate on describing DL transmission for brevity, since similar procedures are followed for UL reception.

\subsection{Zero Forcing (ZF)}

With conventional ZF precoding, as illustrated in Fig.~\ref{fig:threeSchemes}(a), each BS simultaneously serves its scheduled UEs via spatial multiplexing, suppressing all intra-cell interference. However, no inter-cell interference management is performed. Each BS $b$, in a distributed fashion, obtains an estimate of the channel $\mathbf{h}_{ib}$ to each scheduled UE $i \in \mathcal{U}_b$. Neither data or scheduling information is required to be exchanged among BSs. Let
\begin{equation}
\mathbf{H}_b = \left[ {\mathbf{h}}_{1 b},\ldots,{\mathbf{h}_{N_{\mathrm{U},b} b}} \right]
\label{eqn:Hb}
\end{equation}
be the $N_{\mathrm{A}} \times N_{\mathrm{U},b}$ channel matrix at BS $b$, whose columns contain the channel vectors of its scheduled UEs. Then, the ZF precoder
\begin{equation}
\mathbf{W}_b^{\mathrm{ZF}} = \left[ {\mathbf{w}}_{1 b}^{\mathrm{ZF}},\ldots,{\mathbf{w}}_{N_{\mathrm{U},b} b}^{\mathrm{ZF}} \right]	
\end{equation}
at BS $b$ is given by \cite{SpeSwiHaa:04}
\begin{align}
\mathbf{W}_b^{\mathrm{ZF}} = \left(\mathbf{D}_b^{\mathrm{ZF}}\right)^{-\frac{1}{2}} \mathbf{H}_b \left( \mathbf{H}_b^{\mathrm{H}} \mathbf{H}_b \right)^{-1},
\label{eqn:ZF}
\end{align}
where the diagonal matrix $\mathbf{D}_b^{\mathrm{ZF}}$ is chosen to meet the power constraint at each BS with equal UE power allocation, i.e., such that $\Vert \mathbf{w}_{ib}^{\mathrm{ZF}}\Vert^2 = P_{\mathrm{B}}/N_{\mathrm{U},b}$ $\forall i$.

The downlink signal-to-interference-plus-noise ratio (SINR) on a given PRB for UE $i \in \mathcal{U}_b$ is given by
\begin{equation}
\text{SINR}_{i b}^{\mathrm{ZF}} =
\frac{ \left| \mathbf{h}_{ib}^{\mathrm{H}} \mathbf{w}_{ib}^{\mathrm{ZF}} \right|^{2}}
{\sum\limits_{j \in \mathcal{B} \backslash b } \sum\limits_{k \in \mathcal{U}_j} \vert \mathbf{h}_{ij}^{\mathrm{H}} \mathbf{w}_{kj}^{\mathrm{ZF}} \vert^{2} \!+\! I_{i} \!+\! \sigma^{2}_{\epsilon}},
\label{eq:SINR_ZF}
\end{equation}
where $\sigma^{2}_{\epsilon}$ is the variance of the zero-mean complex Gaussian additive thermal noise, and $I_{i}$ is the outdoor co-channel interference received by the UE. Moreover, the intra-cell interference term has been considered negligible owing to both (i) the UE pilot orthogonality during CSI acquisition, and (ii) the high power of the pilot signals received at the BSs in the indoor setup considered. %\mynote{and the intra-cell interference term has been considered negligible under the assumption of perfect channel state information (CSI) acquisition owing to user pilot orthogonality.}

\subsection{Network MIMO (NeMIMO)}

For outdoor deployments, the idea behind network MIMO is to organize BSs in clusters, where BSs lying in the same cluster share information about the data to be transmitted to all UEs in the cluster \cite{HosYuAdv2016}. For the indoor scenario considered in this paper, we assume a single cluster as shown in Fig.~\ref{fig:threeSchemes}(b), i.e., all BSs cooperate to jointly serve the UEs scheduled in all cells \cite{VenLozVal2007}. 
In NeMIMO, all BSs share information about the set of UEs scheduled on each PRB and about the pilot resources assigned to such UEs so as to estimate their channels. More importantly, BSs also share information regarding all data to be transmitted to all scheduled UEs. Let
\begin{equation}
\overline{\mathbf{H}} \triangleq \left[ \mathbf{h}_{1},\ldots, \mathbf{h}_{N_{\mathrm{U}}} \right]	
\label{eqn:Hbar}
\end{equation}
be a $N_{\mathrm{B}} N_{\mathrm{A}} \times N_{\mathrm{U}}$ matrix whose columns denote the channel vector between UE $i$ and all BSs in $\mathcal{B}$, given by $\mathbf{h}_{i} = [{\mathbf{h}}^{\mathrm{T}}_{1 1}, \ldots, {\mathbf{h}}^{\mathrm{T}}_{1 N_{\textrm{B}}}]^{\mathrm{T}}$. Then, the aggregate $N_{\mathrm{B}} N_{\mathrm{A}} \times N_{\mathrm{U}} $ NeMIMO precoding matrix is designed to suppress all crosstalk and it is given by
\begin{align}
\mathbf{W}^{\mathrm{NeMIMO}} = \left(\mathbf{D}^{\mathrm{NeMIMO}}\right)^{-\frac{1}{2}}\overline{\mathbf{H}} \left( \overline{\mathbf{H}}^{\mathrm{H}} \overline{\mathbf{H}} \right)^{-1},
\label{eqn:NeMIMO}
\end{align}
where the diagonal matrix $\mathbf{D}^{\mathrm{NeMIMO}}$ is chosen such that each BS allocates equal power to all UEs, and such that the power constraint is met at every BS, i.e. \cite{BarBocBen2014}
\begin{equation}
\underset{b \in \mathcal{B}}{\max}\left\{ \sum_{n=N_{\mathrm{A}}\left(b-1\right)+1}^{N_{\mathrm{A}} b}  \sum_{i=1}^{N_{\mathrm{U}}} \left| w^{\mathrm{NeMIMO}}_{ni} \right|^2 \right\} = P_{\mathrm{B}}.
\end{equation}
Here, $w^{\mathrm{NeMIMO}}_{ni}$ denotes the entry of $\mathbf{W}^{\mathrm{NeMIMO}}$ on row $n$ and column $i$. Such per-BS power normalization is more fair than assuming a sum-power constraint as in \cite{HosYuAdv2016}, and more practical than solving a complex optimization problem as in \cite{Zha2010}.

% NeMIMO precoding matrix 
%\begin{equation}
%\mathbf{W}^{\mathrm{NeMIMO}} = \left[ \mathbf{w}_{1}^{\mathrm{NeMIMO}},\ldots,\mathbf{w}_{N_{\mathrm{U}}}^{\mathrm{NeMIMO}} \right]	
%\end{equation}

The downlink SINR on a given PRB for UE $i \in \mathcal{U}$, irrespective of its association, is given by
\begin{equation}
\text{SINR}_{i}^{\mathrm{NeMIMO}} =
\frac{ \left| \mathbf{h}_{i}^{\mathrm{H}} \mathbf{w}_{i}^{\mathrm{NeMIMO}} \right|^{2} }
{I_{i} \!+\! \sigma^{2}_{\epsilon}},
\label{eq:SINR_NeMIMO}
\end{equation}
where $\mathbf{w}_{i}^{\mathrm{NeMIMO}} \in \mathbb{C}^{N_{\mathrm{B}} N_{\mathrm{A}}}$ is the $i$-th column of $\mathbf{W}^{\mathrm{NeMIMO}}$ in \eqref{eqn:NeMIMO}, and intra-cell and inter-cell interference terms have been considered negligible as in \eqref{eq:SINR_ZF}.

Although exchanging data information allows BSs to coordinate their transmissions and jointly serve all UEs with an improved SINR, NeMIMO operations come at the cost of severe backhaul requirements in terms of data rate and latency, to enable a tight symbol-level synchronization across multiple BSs. In some cases, said requirements may defy the purpose of a NeMIMO implementation \cite{R117011313,LozHeaAnd:13}.

\subsection{Eigen-direction-aware Zero Forcing (EDA-ZF)}

In this paper, we propose the EDA-ZF precoder as a more practical alternative to NeMIMO to increase the cell-edge throughput via distributed interference mitigation. The reason is that, in contrast with NeMIMO, no data information is shared between BSs. Under EDA-ZF, each BS acquires additional CSI of UEs in neighboring cells, as well as scheduling information from neighboring BSs. As illustrated in Fig.~\ref{fig:threeSchemes}(c), this additional CSI is leveraged by each BS to perform interference suppression towards the channel subspace occupied by neighboring scheduled UEs. While in Section~IV we will show its performance under various scenarios, it should be noted that EDA-ZF is particularly attractive for massive MIMO BS deployments, due to the abundance of spatial degrees of freedom (DoF) provided by large scale antenna arrays. 

During the training phase of EDA-ZF, each BS estimates the channels between itself and all UEs in $\mathcal{U}$ through orthogonal pilots. Let $\mathbf{h}_{k b}$ be such channels, and let $\mathbf{\Sigma}_b$ be a $N_{\mathrm{A}} \times \left( N_{\mathrm{U}} - N_{\mathrm{U}_b} \right)$ matrix whose columns are given by
\begin{equation}
\mathbf{h}_{k b}, \quad k \in \mathcal{U} \backslash \mathcal{U}_b.
\end{equation}
Subsequently, BS $b$ applies a singular value decomposition (SVD) on $\mathbf{\Sigma}_b$, obtaining its singular values sorted in decreasing order ${\nu}_{\ell b}$, $\ell=1,\ldots, \min \left\{ N_{\mathrm{A}}, \left( N_{\mathrm{U}} - N_{\mathrm{U}_b} \right) \right\}$, and its corresponding left singular vectors ${\mathbf{u}}_{\ell b} \in \mathbb{C}^{N_{\mathrm{A}}}$, $\ell=1,\ldots,N_{\mathrm{A}}$. The $N_{\mathrm{N}}$ vectors $\mathbf{u}_{\ell b}$, $k=1,\ldots,N_{\mathrm{N}}$, then span the $N_{\mathrm{N}}$ dominant directions of the channel subspace occupied by the UEs scheduled in neighboring cells. Any power transmitted by BS $b$ on said subspace would generate significant interference at these UEs. For this reason, BS $b$ suppresses the interference generated on the directions $\mathbf{u}_{\ell b}$, $\ell=1,\ldots,N_{\mathrm{N}}$ during data transmission.\footnote{It is more fair to suppress interference on the eigendirections of the aggregate channel than on the channel directions of certain UEs. In fact, when there are less DoF for nulls than neighboring UEs, the latter approach relieves certain UEs of all interference while not suppressing any to others.} This is accomplished by sacrificing $N_{\mathrm{N}}$ spatial DoF to place radiation nulls, as illustrated by the red arrows in Fig.~\ref{fig:threeSchemes}(c). Let
\begin{equation}
\widetilde{\mathbf{H}}_b \triangleq \left[ \mathbf{h}_{1 b},\ldots,{\mathbf{h}}_{N_{\mathrm{U},b} b}, \mathbf{u}_{1 b} \ldots, \mathbf{u}_{N_{\mathrm{N}} b} \right],	
\label{eqn:Htilde}
\end{equation}
be a $N_{\mathrm{A}} \times \left( N_{\mathrm{U},b} + N_{\mathrm{N}} \right)$ matrix whose columns contain the channels vectors of all UEs scheduled by BS $b$, as well as the spatial directions $\mathbf{u}_{\ell b}$ to null, $\ell=1,\ldots,N_{\mathrm{N}}$. Then, the EDA-ZF precoder $\mathbf{W}_b^{\mathrm{EDA-ZF}}$ 
at BS $b$ is given by the first $N_{\mathrm{U},b}$ columns of the matrix $\widetilde{\mathbf{W}}_b^{\mathrm{EDA-ZF}}$ defined as
\begin{align}
\widetilde{\mathbf{W}}_b^{\mathrm{EDA-ZF}} = \left( \mathbf{D}_b^{\mathrm{EDA-ZF}}\right)^{-\frac{1}{2}} \widetilde{\mathbf{H}}_b \left( \widetilde{\mathbf{H}}_b^{\mathrm{H}} \widetilde{\mathbf{H}}_b \right)^{-1},
\label{eqn:EDA-ZF}
\end{align}
where the diagonal matrix $\mathbf{D}_b^{\mathrm{EDA-ZF}}$ is chosen for equal UE power allocation, i.e., such that $\Vert \mathbf{w}_{ib}^{\mathrm{EDA-ZF}}\Vert^2 = P_{\mathrm{B}}/N_{\mathrm{U},b}$ $\forall i$, with $\mathbf{w}_{kj}^{\mathrm{EDA-ZF}}$ denoting the $k$-th column of \eqref{eqn:EDA-ZF}.

The downlink SINR on a given PRB for UE $i$ served by BS $b$ is given by
\begin{equation}
\text{SINR}_{i b}^{\mathrm{EDA-ZF}} =
\frac{ \left| \mathbf{h}_{ib}^{\mathrm{H}} \mathbf{w}_{ib}^{\mathrm{EDA-ZF}} \right|^{2}}
{\sum\limits_{j \in \mathcal{B} \backslash b } \sum\limits_{k \in \mathcal{U}_j} \left| \mathbf{h}_{ij}^{\mathrm{H}} \mathbf{w}_{kj}^{\mathrm{EDA-ZF}} \right|^{2} \!+\! I_{i} \!+\! \sigma^{2}_{\epsilon}},
\label{eq:SINR_EDA-ZF}
\end{equation}
where, similarly to \eqref{eq:SINR_ZF} and \eqref{eq:SINR_NeMIMO}, the intra-cell interference term has been neglected.

%% file: Section4.tex
\section{Performance Evaluation}

In this section, we compare the DL performance of the three interference management schemes described in Section~III and depicted in Fig.~\ref{fig:threeSchemes}. We investigate three deployment scenarios with different BS densities: (i) a \emph{sparse} deployment of $N_{\mathrm{B}}=2$ massive MIMO BSs with $N_{\mathrm{A}}=64$ antennas each; (ii) an \emph{intermediate} deployment of $N_{\mathrm{B}}=8$ BSs with $N_{\mathrm{A}}=16$ antennas each; and (iii) a \emph{dense} deployment of $N_{\mathrm{B}}=32$ BSs with $N_{\mathrm{A}}=4$ antennas each.\footnote{Deploying two BSs (resp. one BS) yields a minimum received signal strength indicator (RSSI) \cite{3GPP36214} of $-$89.2 dBm (resp. $-$98.0 dBm) at the edges of the scenario. This accounts for the transmit power, the antenna pattern, and the path loss. Since the UE sensitivity is $-$94 dBm when operating in TDD \cite{3GPP36101}, we do not consider deploying less than two BSs to guarantee coverage.} 
In the sparse case, BSs are deployed as in Fig.~\ref{fig:threeSchemes}. In the intermediate and dense cases, BSs are uniformly deployed following regular $2\times 4$ and $4\times 8$ grids, respectively \cite{3GPP38802}. We keep the total number of BS antennas $N_{\mathrm{B}} N_{\mathrm{A}} = 128$ constant in all three scenarios, in order to observe the effect of densification. For a fair comparison, we also fix the total power as $P_{\mathrm{tot}}$, and set the power per BS as $P_{\mathrm{B}} = P_{\mathrm{tot}} / N_{\mathrm{B}}$. We assume that each BS schedules a maximum number of UEs for multi-user DL transmission, i.e., $N_{\mathrm{U}_b} \leq N_{\mathrm{A}}/4$, $\forall b$. In practice, the allocation of DoF could be dynamically optimized by trading multiplexing gain for beamforming and nulling capabilities. We also consider link adaptation, where for each SINR value the modulation and coding scheme (MCS) is selected to ensure a block error rate (BLER) of $10^{-1}$, and the maximum MCS is 256-QAM with code rate of 0.93 \cite{3GPP36213}. Note that while the theoretically maximum MCS rates are 86.3 Mbps, those plotted in the following account for the fraction of time each UE is scheduled. This is because more UEs are deployed than those that can be spatially multiplexed simultaneously, and therefore they must also be multiplexed in time. Table~\ref{table:deployment} summarizes the three deployment scenarios, whereas a detailed list of all system parameters is provided in Table~\ref{table:parameters}.

\vspace*{-0.04cm}

\subsection{Deployment Densification and Indoor Interference}

\begin{table}
\centering
\caption{Deployment scenarios and parameters}
%\vspace{-0.3cm}
\label{table:deployment}
\def\arraystretch{1.1}
\begin{tabulary} {\columnwidth} {|L|C|C|C|}
\hline
\textbf{Parameter/scenario} 					& \textbf{sparse} & \textbf{intermediate} & \textbf{dense}	\\ \hline
Number of BSs, $N_{\mathrm{B}}$				     & 2 & 8 & 32			\\ \hline
Antennas per BS, $N_{\mathrm{A}}$			     & 64 & 16 & 4  		\\ \hline
Max. scheduled UEs per BS & 16 & 4 & 1 							\\ \hline

\end{tabulary}
\end{table}

\begin{table}
\centering
\caption{System parameters}
%\vspace{-0.3cm}
\label{table:parameters}
\def\arraystretch{1.1}
\begin{tabulary}{\columnwidth}{ |p{3.4cm} | p{3.9cm} | }
\hline
\textbf{Parameter} 					& \textbf{Description} 		\\ \hline
BS transmit power, $P_\mathrm{B}$ 			& Sparse: $24$ dBm \cite{3GPP38802}; intermediate: $18$ dBm; dense: $12$ dBm.								\\ \hline
BS antenna array         	& Uniform square planar array						\\ \hline
BS antenna elements         & $5$ dBi with $90^{\circ}$ half-power beam width \cite{3GPP38802}						\\ \hline
UE antenna elements         & Omnidirectional with 0 dBi \cite{3GPP38802}						\\ \hline
%System bandwidth 					& 20 MHz \cite{3GPP38802}															\\ \hline
Carrier frequency; bandwidth					& 4 GHz \cite{3GPP38802}; 20 MHz \cite{3GPP38802}			\\ \hline
UE noise figure 					& 9 dB \cite{3GPP38802} 		\\ \hline
Path loss and prob. of LoS 				& InH \cite{3GPP38802} 		\\ \hline
Shadowing 										& Log-normal with $\sigma = 3/4 $ dB (LoS/NLoS) \cite{3GPP38802} 			\\ \hline
Fast fading 									& Ricean with log-normal K factor \cite{3GPP38802} and Rayleigh multipath	\\ \hline
Thermal noise 									& $-$174 dBm/Hz spectral density	\\ \hline
Outdoor interference								& Ranging from no interference to $-60$ dBm over 20 MHz	\\ \hline
Floor size									& $120~\textrm{m}\times 50~\textrm{m}$ \cite{3GPP38802}		\\ \hline
%BS positions								& Ceiling mounted, equally spaced in central corridor as in Fig.~\ref{fig:threeSchemes}			\\ \hline
BS and UE heights								& 3 and 1.5 meters\cite{3GPP38802}			\\ \hline
UE distribution 								& 80 uniformly deployed UEs 	\\ \hline
%BS scheduling 								& Round Robin scheduler	\\ \hline
UE association; scheduling 						& RSS-based; Round Robin	\\ \hline
Link adaptation						& MCS selected for BLER = $10^{-1}$. Maximum MCS: 256-QAM with code rate 0.93 \cite{3GPP36213}.	\\ \hline
\end{tabulary}
\end{table}

\begin{figure}[!t]
\centering
\includegraphics[width=\figwidth]{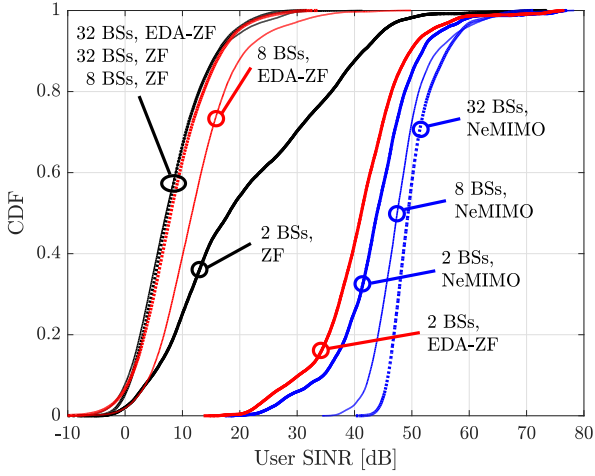} %
\caption{UE SINR per PRB for ZF, NeMIMO, and EDA-ZF, under sparse, intermediate, and dense deployments.}
\label{fig:SINR_noInt}
\end{figure}

\begin{figure}[!t]
\centering
\includegraphics[width=\figwidth]{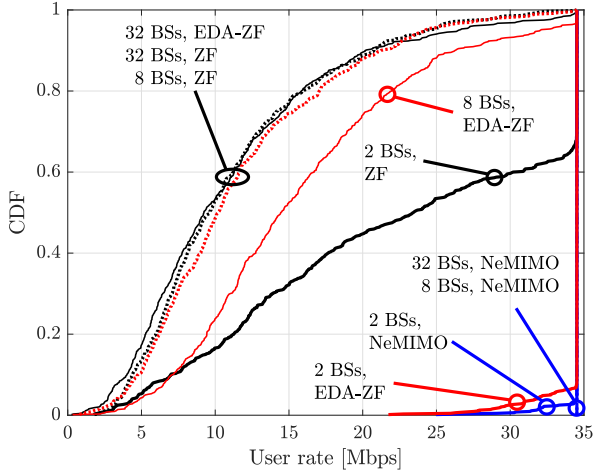} %
\caption{UE DL rate for ZF, NeMIMO, and EDA-ZF, under sparse, intermediate, and dense deployments.}
\label{fig:rates_noInt}
\end{figure}

Fig.~\ref{fig:SINR_noInt} and Fig.~\ref{fig:rates_noInt} compare the performance of the three interference management schemes under sparse, intermediate, and dense deployments without co-channel outdoor interference, i.e., by forcing $I_i=0$ $\forall i$. For EDA-ZF, $N_{\mathrm{N}}=N_{\mathrm{A}}/4$ radiation nulls are placed by each BS. The two figures respectively show the cumulative distribution function (CDF) of the user DL SINR per PRB and of the DL rate. The following two observations can be made from these figures.
\subsubsection{Interference suppression}
Under conventional ZF the 5\%-worst performance is significantly lower than the average one, showing that cell-edge UEs -- affected by strong inter-cell interference -- are heavily penalized. On the contrary, implementing the proposed EDA-ZF scheme in a sparse deployment with two massive MIMO BSs provides uniformly good performance for all UEs. In particular, the achievable rates are very close to the ones supported by the maximum MCS, and similar to the ones attained by NeMIMO, which unlike EDA-ZF requires full coordination among BSs.
\subsubsection{Densification}
Densifying the deployment worsens the performance of both ZF and EDA-ZF. This is due to the large number of LoS links, which makes the indoor scenario strongly interference limited. In other words, the damage of a larger inter-cell interference outweighs the benefit of an increased proximity to UEs \cite{LopDinCla:15}. On the other hand, NeMIMO benefits from densification, since BSs gain proximity to UEs while remaining devoid of inter-cell interference. Densification is thus detrimental unless full coordination is implemented.

\vspace*{-0.04cm}

\subsection{Impact of Co-channel Outdoor Interference}

\begin{figure}[!t]
\centering
\includegraphics[width=\figwidth]{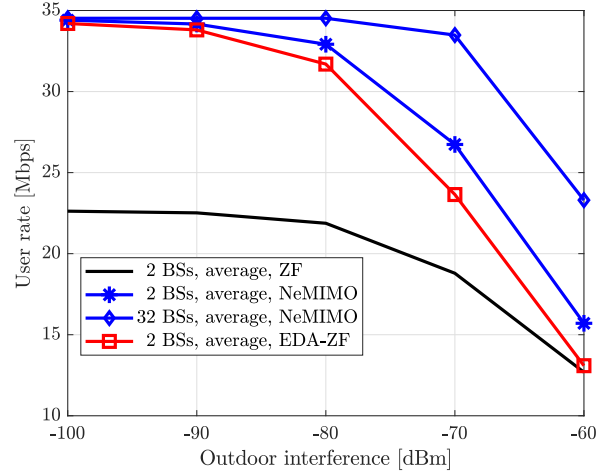} %
\caption{UE DL average rate versus co-channel outdoor interference for ZF, NeMIMO, and EDA-ZF.}
\label{fig:rates_avg_vsInt}
\end{figure} 

%\vspace*{-0.3cm}

\begin{figure}[!t]
\centering
\includegraphics[width=\figwidth]{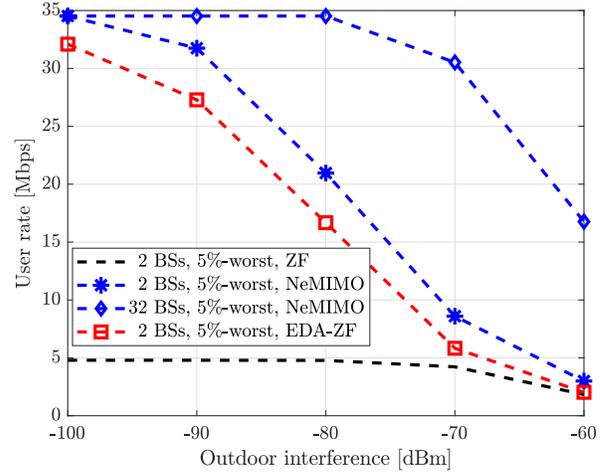} %
\caption{UE DL 5\%-worst rate versus co-channel outdoor interference for ZF, NeMIMO, and EDA-ZF.}
\label{fig:rates_five_vsInt}
\end{figure} 

Fig.~\ref{fig:rates_avg_vsInt} and Fig.~\ref{fig:rates_five_vsInt} evaluate the effect of co-channel outdoor interference -- denoted $I_i$ in \eqref{eq:SINR_ZF}, \eqref{eq:SINR_NeMIMO}, and \eqref{eq:SINR_EDA-ZF} -- on the average and 5\%-worst UE rates, respectively. Following \cite{NguWigKov2017}, we consider values of $I_i$ ranging between $-$100~dBm and $-$60~dBm over the 20~MHz bandwidth. The highest value may be generated, e.g., by a macro BS located 200 meters away, transmitting 43~dBm over a directional antenna of 17~dBi pointing towards the building of interest. Lower values of interference could be experienced if the interfering macro BS is farther away, if the building is shadowed by other buildings, or if the outer wall material and thickness cause a higher penetration loss \cite{NguWigKov2017}. The figures show that for lower values of the outdoor interference, deploying two massive MIMO indoor BSs and operating them according to the proposed EDA-ZF is almost optimal in terms of both average and 5\%-worst performance, since it approaches the maximum MCS rates. Moreover, it can be observed that EDA-ZF approaches the performance of NeMIMO in the sparse deployment scenario irrespective of the outdoor interference, thanks to the interference suppression capabilities of the massive antenna arrays. For higher values of the outdoor interference, a dense deployment of NeMIMO BSs brings transmitters closer to receivers, without introducing additional indoor interference. This especially rewards cell-edge UEs, as observed in the increased 5\%-worst rate, but at the expense of demanding full coordination among BSs.

%\begin{figure}[!t]
%\centering
%\includegraphics[width=\columnwidth]{Figures/rates_vsInt.eps} %
%\caption{UE DL average and 5\%-worst rate vs. co-channel outdoor interference for ZF, NeMIMO, and EDA-ZF.}
%\label{fig:rates_vsInt}
%\end{figure} 

\subsection{Degrees of Freedom Allocation for Nulls}

\begin{figure}[!t]
\centering
\includegraphics[width=\columnwidth]{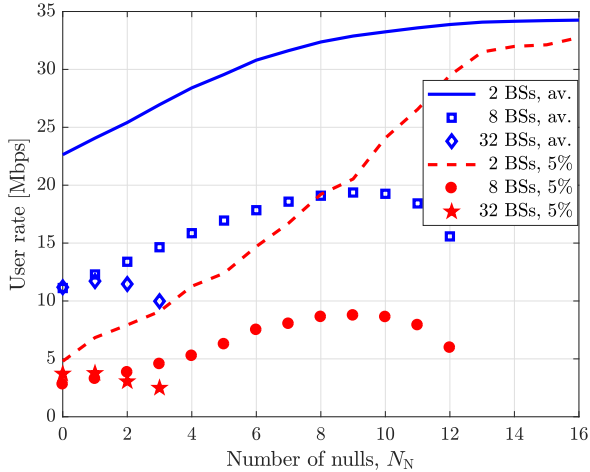} %
\caption{UE DL average and 5\%-worst rate for EDA-ZF versus number of DoF employed for nulls, for sparse, intermediate, and dense deployments.}
\label{fig:rates_vsNulls}
\end{figure} 

Fig.~\ref{fig:rates_vsNulls} shows the performance of EDA-ZF as a function of the number of DoF allocated for nulls. Average and 5\%-worst rates are plotted for the sparse, intermediate, and dense deployment cases. With two massive MIMO BSs, each BS is equipped with $N_{\mathrm{A}}=64$ antennas and serves a maximum of $16$ UEs. In this case, it pays off allocating the maximum number of DoF for nulls, i.e., $N_{\mathrm{N}}=16$, to suppress all inter-cell interference. The gain is especially noticeable for the $5\%$-worst rate, which increases by seven-fold when varying $N_{\mathrm{N}}$ between 0 and 16. With intermediate and dense deployments, BSs cannot suppress all interference due to their reduced number of antennas. A tradeoff thus exists between interference suppression (more nulls) and beamforming gain (fewer nulls). %In the scenarios we considered, both average and $5\%$-worst rates are maximized by roughly allocating one quarter of the available DoF for nulls, i.e., $N_{\mathrm{N}} = N_{\mathrm{A}}/4$.

%% file: Section5.tex
\section{Conclusion}

We tackled the issue of indoor inter-cell interference management with the aim of providing \emph{uniformly high wireless capacity}. To this end, we proposed eigen-direction-aware ZF for distributed inter-cell interference mitigation, and we compared its performance against network MIMO -- which targets a complete inter-cell interference removal -- and conventional ZF without BS inter-cell interference management. Our results demonstrated that \emph{indoor massive MIMO deployments}, paired with the proposed eigen-direction-aware ZF, approach the data rates achieved by network MIMO for all users in the network. The proposed scheme does not require a tight BS coordination with symbol-level synchronization nor full data exchange as in network MIMO, therefore providing a compelling alternative for future high-capacity wireless indoor networks.